\documentstyle[12pt,epsf]{l-aa}

\def\M{{\cal M}}
\def\R{{\cal R}}

\def\beq#1{\begin{equation}\label{#1}}
\def\eeq{\end{equation}}
\def\beqa#1{\begin{eqnarray}\label{#1}}
\def\eeqa{\end{eqnarray}}

\def\myfrac#1#2{\left(\frac{#1}{#2}\right)}
\def\comment#1{\relax}

\def\spose#1{\hbox to 0pt{#1\hss}}
\def\simlt{\mathrel{\spose{\lower 3pt\hbox{$\mathchar"218$}}
     \raise 2.0pt\hbox{$\mathchar"13C$}}}
\def\simgt{\mathrel{\spose{\lower 3pt\hbox{$\mathchar"218$}}
     \raise 2.0pt\hbox{$\mathchar"13E$}}}
\def\simpropto{\mathrel{\spose{\lower 3pt\hbox{$\mathchar"218$}}
     \raise 2.0pt\hbox{$\propto$}}}

\makeatletter
\def\eqalign#1{\null\,\vcenter{\openup\jot\m@th
  \ialign{\strut\hfil$\displaystyle{##}$&$\displaystyle{{}##}$\hfil
      \crcr#1\crcr}}\,}
\def\eqalignleft#1{\null\,\vcenter{\openup\jot\m@th
  \ialign{\strut$\displaystyle{##}$\hfil&$\displaystyle{{}##}$\hfil
      \crcr#1\crcr}}\,}
\makeatother
\def\M{{\cal M}}

\def\R{{\cal R}}

\begin{document}
\setcounter{footnote}{1}

\thesaurus{12       % Physical and Chemical Processes
     (02.07.2; % {Gravitational waves
      08.02.1)} % Stars: binaries: close
     
\title{On the Gravitational Wave Noise from Unresolved 
Extragalactic Binaries}
\author{
D.I.Kosenko
K.A.~Postnov}

\institute{
Faculty of Physics and 
Sternberg Astronomical Institute, Moscow University,
                119899 Moscow, Russia}
\date{Received ... 1998, accepted ..., 1998}
\maketitle
\markboth{D.Kosenko \& K.Postnov. Extragalactic Binary GW Noise }{ ...}

\begin{abstract}
We calculate stochastic gravitational wave background produced by
extragalactic merging binary white dwarfs at the LISA frequencies
$10^{-3}-10^{-2}$ Hz with account of a strong evolution of global star
formation rate in the Universe recently established observationally. We show
that for the observed global star formation history and modern cosmological
models the extragalactic background is an order of magnitude smaller than
the mean Galactic value. An early star formation burst at high redshifts can
bring it at a higher level but still a few times lower than the mean
Galactic one.


\end{abstract}

\section{Introduction}

Gravitational waves (GW) from unresolved binary stars in the Galaxy forms a
confusion limit in the frequency domain of LISA laser space interferometer
(Bender et al. 1996)
$10^{-4}-10^{-1}$ Hz that effectively adds to the detector's noise.
Different aspects of this background have been studied earlier by Rosi and
Zimmerman (1976), Lipunov
\& Postnov (1987), Lipunov, Postnov \& Prokhorov (1987), Hils, Bender \&
Webbink (1990), Bender \& Hils (1997), Giampieri \& Polnarev (1997),
Postnov \& Prokhorov (1998).

The importance of the knowledge of binary confusion limit within the LISA
frequency band is dictated by many factors -- for example, it limits the
angular resolution of the detector (Cutler 1997) and restricts possibilities
of observing possible relic GW backgrounds (see, e.g., discussion in
Grishchuk 1997). Since LISA with its three arms is expected to operate as
essentially single interferometer (Schutz 1997), it can only see noise
sources above its intrinsic noise $h_{rms}$. In our previous paper (Postnov
\& Prokhorov 1998) we focused on the stochastic GW noise from galactic
binaries and showed that, depending on (unknown) galactic binary white dwarf
merger rate, this noise becomes lower than the expected LISA rms noise at
frequencies $f\sim 3\times 10^{-2}$ Hz \footnote{This limiting frequency is
derived assuming integration time $T$ and frequency $f$ such that $fT=1$;
for a fixed integration time (say, 1 year) the frequency resolution is
$\Delta f\approx 1/T\sim 10^{-7}$ Hz. Correspondingly, the limiting
frequency for the galactic white dwarf binary confusion limit would decrease
as $(fT)^{-3/8}$, typically by 50 times at 1 mHz for 1 year integration
time.} We concluded that to be detectable by LISA, relic GW backgrounds
should be as high as
$\Omega_{GW}h_{100}^2>10^{-8}$ at $10^{-2}$ Hz.

Clearly, the galactic binary GW noise, tracing the distribution of stars in
the Galaxy, should be highly anisotropic (of order of one magnitude higher
in the direction of the Galactic center; see calculations by Lipunov et al.
1995) and thus modulated with LISA turning in its orbit. This distincitve
feature of the galactic stochastic background can in fact be used to measure
it by space-born interferometers (Giampieri \& Polnarev 1997). This is not
the case for a GW-background produced by extragalactic binaries, which
becomes important at frequencies where galactic binaries do not form
continuous noise ($f> 3\times 10^{-3}$ Hz for 1-year integration).

A crude estimate shows that the isotropic extragalactic background is
expected to be an order of magnitude smaller than the average GW noise from
galactic binaries (see Lipunov et al. (1987, 1995), Hils et al. (1990)). All
estimates of this background made so far, however, have not taken into account
the fact of a strong global star formation evolution recently revealed by
different 
astronomical observations (see the results of Canada-France-Hawaii survey of
far galaxies with $z\simlt 1$ (Lilly et al. 1996), the analysis of galaxies
with Lyman jump at $z\simgt 2$ (Madau et al. 1996), and the analysis of the
HST deep field survey of galaxies with $1<z<2$ (Connoly et al. 1997)).
According to these studies, the global star formation rate strongly
increases with redshift, peacking at $z\sim 2$ (see Fig. 1 below reproduced
from Connoly et al. (1997), with points from Madau et al. (1996) corrected
for extinction as in Boyle \& Terlevich (1997)). 

In a Euclidean space, the remote sources would contribute more and more to
the energy density with distance $r$, as $\propto r$, leading to the
photometric paradox; in a FRW expanding Universe this is not the case.
However, the effects of proper density evolution of galaxies (star
formation) with redshift, together with the proper evolution of the sources
inside galaxies, could compensate for the cosmological energy density
dilation. So a priori it is not at all clear if the remote extragalactic
binaries can be neglected in studying extragalactic binary GW background.
This is the motivation to our calculations. 

%We have found that,
%depending on the redshift of the initial star formation epoch 
%$z_*$, extragalactic
%binary GW background could be comparable with the average galactic level.

\section{GW noise from sources with changing frequency}

We start with a short description of how a GW background is formed by a
population of frequency-unresolved sources with changing frequency. The
amplitude of the stochastic GW background is conventionally measured in
terms of the energy density per logarithmic frequency interval related to
the critical density to close the Universe

\beq{Omega_GW:def}
\Omega_{GW}=\frac{dE_{GW}}{d\ln f\,d^3x\rho_{cr}}\,,
\eeq
where $\rho_{cr}=3H_0^2/8\pi \approx 1.9\times 10^{-29}
h_{100}^2$ g cm$^{-3}$, 
$h_{100}=H_0/100$ [km/s/Mpc]
is the present value of the Hubble constant; we shall use geometrical units
$G=c=1$ throughout the paper. As in the case of deterministic GW signals,
one may equally use the characteristic dimensionless 
amplitude of metric variations
$h_c$, which determines the signal-to-noise ratio
on the detector 
$(S/N)=h_c/h_{rms}$. This amplitude is connected with 
$\Omega_{GW}$ through
%\beq{hc_stoch:def}
$
h_c(f)=(1/2\pi)(H_0/f)\Omega_{GW}^{1/2}\,.
$
%\eeq

To calculate GW noise produced by some frequency-unresolved sources, we need
know the number of sources per logarithmic frequency interval. At the first
glance, this would require knowing the precise formation and evolution of
sources. However, when only GW carries away the angular momentum from the
emitting source (as is the case for coalescing binary white dwarf forming
the high-frequency part of the galactic binary GW background), the problem
becomes very simple and physically clear.

Energy loss causes the change in the GW frequency $\omega=2\pi f$ of emitting
objects. In the case of a binary star with masses of the components $M_1$
and $M_2$ in a circular orbit of radius
$a$ the negative orbital energy, $E_{orb}= -M_1M_2/2a\sim \M f_{orb}^{2/3}$
(here $\M=(M_1+M_2)^{2/5}(M_1M_2)^{3/5}$ is the so-called ``chirp mass'' of
the system) is being lost, and the orbital frequency (hence, the GW
frequency, which is twice the orbital one) increases.

To a good approximation, the conditions of star formation and
evolution of astrophysical objects in our Galaxy may be viewed as
stationary. This is true at least for last 5 billion years.  Let the
formation rate of GW sources be $\R$. 
Inside the LISA frequency range, $10^{-4}-10^{-1}$ Hz, only coalescing
binary white dwarfs (WD) and binary neutron stars contribute. Even if
binary neutron stars coalesce at a rate of 1/10000 yr in the Galaxy,
their total number at any moment of time 
still should be much smaller than of the WD
binaries, and we restrict ourselves to considering
only binary WD. As we showed in our previous paper 
(Postnov \& Prokhorov 1998), the galactic rate of binary WD
mergers may be restricted by the statistics of 
SN type Ia explosions, which are much more numerous 
than the observed close binary WDs. However, as we
show below, the ratio of the extragalactic to galactic 
GW binary noise is almost independent of the exact value of
$\R$. In real galaxies $\R$ may be the function of time,
which will be accounted for by the star formation history 
SFR($t$).

At some moment of time
the number of sources per unit
logarithmic frequency interval is
\begin{equation}
dN(t)/d\ln f \equiv N(f,t) = \R(t) (f/\dot f)\,.
\label{contin}
\end{equation}
so the total energy emitted in GW per second per unit logarithmic
frequency interval at frequency 
$f$ by all such sources in the galaxy is
\begin{equation}
\eqalignleft{
dE_{GW}/(dt\,d\ln f) &\equiv L(f)_{GW}= \widetilde L(f)_{GW} N(f,t)\cr
&=\widetilde L(f)_{GW} \R(t) \times (f/\dot f)\,,
\cr}
\end{equation}
where $\widetilde L(f)_{GW}$ is the GW luminosity (erg/s) of the typical
source at frequency $f$, $\widetilde L(f)_{GW}\propto f^{10/3}$
for binary systems.

For an isotropic background we find
\begin{equation}
\Omega_{GW}(f)\rho_{cr}  = L(f)_{GW}/(4\pi  \langle r \rangle^2)
\label{omega}
\end{equation}

where $\langle r \rangle$ is the inverse-square average distance to
the typical source.  Strictly speaking, this distance (as well as the
binary chirp mass $\cal M$) may be a
function of frequency since the binaries characterized by different
$\cal M$ may be differently distributed in the Galaxy. 
Since the real distribution of binaries of different mass in the
Galaxy is poorly known, we can take, as an estimate,
the mean photometric distance for a spheroidal
distribution in the form
$dN\propto \exp[-r/r_0]\,\exp[-(z/z_0)^2]$ 
($r$ is the radial
distance to the Galactic center and $z$ the hight above the galactic
plane) with $r_0=5$ kpc and $z_0=4.2$ kpc with $\langle r \rangle
\approx 7.89$ kpc.  

For binary stars the energy reservoir radiated in GW is 
their orbital binding energy,
which is  a power law of the orbital frequency: $E\propto
f^{2/3}$. Hence the frequecy
change $(f/\dot f)$ may be found from the equation
$
dE/dt=(2/3) E (\dot f/f)
$
By energy conservation law 
$
dE/dt=(dE/dt)_{GW}+(dE/dt)_{\ldots}
$
where ($\ldots$) means
other possible loss/gain of energy. Finally, we obtain
\begin{equation}
(f/\dot f)= (2/3) E / ((dE/dt)_{GW}+(dE/dt)_{\ldots})
\end{equation}
and 
\begin{equation}
L(f)_{GW}= (2/3) \R  \widetilde E \frac{1}{
1+\frac{(dE/dt)_{\ldots}}{(dE/dt)_{GW}}}
\end{equation}
Remarkably, if GW is the dominant source of energy
removal, the resulting GW stochastic background depends only on the
source formation rate:
\begin{equation}
\Omega_{GW}(f)\rho_{cr}  = (2/3)\R  \widetilde E /(4\pi  \langle
r \rangle^2)
\label{omega_R}
\end{equation}

For the frequency range under consideration the assumption that only GW
emission drives the orbital evolution of a close binary WD is a perfect
approximation. The same relates to the assumption tacitly made that the
orbit of all such binaries is circular. As we are interested in the blue end
of this frequency range, we need not to consider much more complicated
evolution of cataclysmic binaries whose orbital periods are always higher
than a few tens of minutes.

\section{GW noise from unresolved extragalactic binary stars at LISA
frequencies}

Let the detector's frequency be $f$. Then sources in galaxies within the
comoving volume element $dV(z)$ at redshift $z$ produce an isotropic GW
background with the energy densidy near the detector (at $z=0$)
\beq{dO}
\rho_{cr} d\Omega_{GW}(f)=\myfrac{L(f')}{4\pi D(z)^2}n(z) dV(z) \,,
\eeq
where $L(f)$ is the GW luminosity 
of one galaxy at the comoving frequency $f'=f(1+z)$,
$D(z)$ is the photometric distance,
$n(z)=n_GF(z)(1+z)^3$ is the comoving density of galaxies 
at redshift $z$ with factor $F(z)$ describing the galactic density
evolution, and  
$n_G$ is the present day density of galaxies. In the case of no 
proper evolution $F(z)=1$.

Transiting in a standard way from integration over proper volume to
integration over  cosmic time (e.g. Weinberg 1972) we arrive at
\beq{dOz}
\rho_{cr} \Omega_{GW}(z^*)=n_G\int_0^{z^*}F(z)
\myfrac{L(f(1+z))}{1+z}\frac{dt}{dz}dz \,,
\eeq
where $z^*$ marks the beginning of the star formation in the Universe. 
Noticing that $L(f')\propto E_{orb}(f(1+z))$ scales  with redshift as 
$(1+z)^{2/3}$, using  Eq. 
(\ref{dOz}), and assuming that binary WD merging rate $\R$
per proper volume changes with time only due to galactic density evolution 
$F(z)$, we finally obtain the ratio of extragalactic to galactic
GW energy density produced by coalescing binary WD:
\beq{ratio}
\frac{\Omega_{GW}(z^*)}{\Omega_{GW}}=
4\pi \langle r^2\rangle n_G\int_0^{z^*}\frac{F(z)}{(1+z)^{-1/3}}\frac{dt}{dz}dz
\eeq

The present-day galactic density can be rewritten 
in terms of $\Omega_b$, the baryon fraction in stars
relative to the critical density, assuming the stellar mass
of the typical galaxy to be  
$M_G=10^{11} M_\odot$:
\beq{n_G}
n_G \approx 0.013( \hbox{Mpc}^{-3})\myfrac{\Omega_b}{0.005}h_{100}^2\,.
\eeq
Here $\Omega_b$ is normalized to the observed 
present-day  baryon density of luminous matter  
($\Omega_b=0.003$ inside bulges of spiral and in elliptical
galaxies, $\Omega_b=0.0015$ in disks of spiral galaxies;
see Fukugita et al 1996, 1997 for further details). For 
$h_{100}=0.7$ the numerical coefficient in (\ref{n_G})
is in a good agreement with the measured local galactic density
$0.0048 \hbox{Mpc}^{-3}$
(Loveday et al. 1992).

The time-redshift relation in a FRW cosmological model with 
arbitrary matter ($\Omega_M$) and vacuum ($\Omega_\Lambda$) density
is
\beq{t(z)}
dt = \frac{1}{H_0} \frac{dz}{(1 + z)\sqrt{(1+z)^2(1 + \Omega_M z) - z(2 +
z) \Omega_{\Lambda}}}
\eeq
(notice that $\Omega_M$ includes 
both visual baryon density $\Omega_b$ and dark matter).

For a flat ($\Omega_M=1$) FRW universe 
without cosmological term 
($\Omega_\Lambda=0$) $dt/dz=(1+z)^{-5/2}$ and without source evolution
($F(z)=1$) 
the integrand in Eq. (\ref{ratio}) turns into $(1+z)^{-17/6}$, so
that\footnote{Eq. (14) in paper (Postnov \& Prokhorov (1998)) was
calculated for $F(z)=(1+z)^3$, so it should be substituted
by this equation for non-evolving sources.} 
\beq{ratio_flat}
\eqalignleft{
&\frac{\Omega_{GW}(z^*)}{\Omega_{GW}}\approx \cr
&0.03\myfrac{\Omega_b}{0.005}
h_{100}\myfrac{\langle r\rangle}{10 \hbox{kpc}}^2[1-(1+z^*)^{-11/6}]\,.
\cr}
\eeq  
Clearly, the extragalactic background is small in comparison with the
galactic one if no cosmological evolution of global star formation rate is
included.

Now we wish to take into account the evolution of WD merging rate with time
(redshift). Let $G(t)$ be the time dependence of the event rate of interest
after a $\delta$-function-like star fromation burst. This function can be
considered as a Green function for arbitrary law of star formation SFR(t).
Then the comoving event rate at any time can be calculated as
\beq{SFR} 
\R_G(t)=\int\limits_{t(z_*)}^t{\rm SFR}(\tau)G(t-\tau)d\tau 
\eeq
(here $t(z_*)$ is the initial moment of star formation).
The event rate per proper time observed from a layer $dz$  at
redshift $z$ is
\beq{R(z)}
\R(z)=n(z)\R_G(z)dV(z)\,.
\eeq

Clearly, the evolution of the comoving galactic density $n(t(z))$ can be
considered in the same way, but since we derive from observations the
comoving luminosity density as a function of redshift, it is impossible to
separate the star formation rate inside galaxies and the comoving galactic
density evolution without making special model assumptions. We will assume
that expressing the comoving density of galaxies through the luminous baryon
density $\Omega_b$ (\ref{n_G}) reflects real global SFR evolution without
knowing how precisely has the comoving density of different types of
galaxies changed with time. This remains valid until $\Omega_b$ is assumed
constant. For example, $\Omega_b$ may change in models with non-zero
cosmological constant at stages when its influence on the expansion of the
Universe becomes dominant.

\begin{figure} 
\centerline{ 
\epsfxsize=0.6\hsize
%\epsfbox{../../diss/figures/stoch/sfrnew.eps}} \caption{The evolution 
\epsfbox{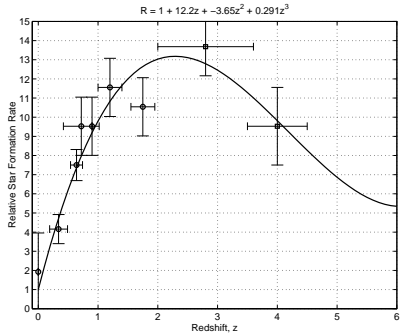}} \caption{The evolution 
of global star formation rate in the Universe from 
Connolly et al (1997) normalized such that the present
star formation rate in solar masses per year per 
a $10^{11} M_\odot$ galaxy is unity. Open squares 
at large redshifts represent data from Madau et al. (1996)
corrected for extinction as in 
Boyle and Terlevich (1997).}
\label{f:SFR} 
\end{figure}

The observed star formation rate evolution SFR(z) were taken from 
Connolly et al. (1997). In fact, the absolute value of 
star formation rate is deduced from the observed UV luminosity 
density, which requires knowledge of the initial mass function 
of stars and models of stellar evolution (Madau et al. 1997).  
For the relative star formation these requirements become
weaker (at least so far as we assume the same parameters 
of the star formation and evolution at any time). 
We reproduce this function 
(normalized such that SFR$(z=0)=1$) in Fig. 
\ref{f:SFR} with data from 
Madau et al. (1996) (open squares) 
being corrected for extinction 
as in Boyle and Terlevich (1997)
(then the global SFR coincides with the evolution 
of the comoving density of quasars). For numerical calculations
this SFR can be approximated by the polynomial 
\beq{SFR(z)} 
{\rm SFR}(z)\approx 1+12.2z-3.65z^2+0.291 z^3 \,.
\eeq

\begin{figure}
\centerline{  
\epsfxsize=0.8\hsize
%\epsfbox{../../diss/figures/stoch/rate.eps}}
\epsfbox{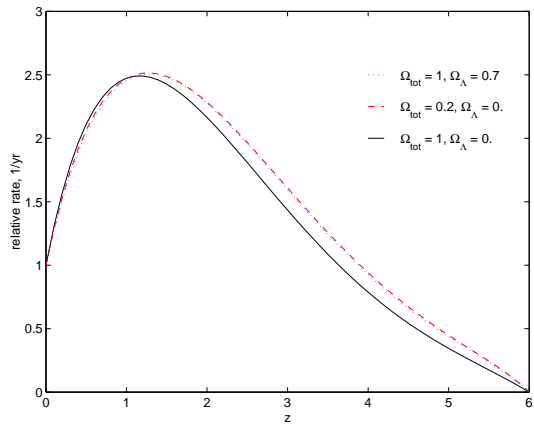}}
\caption{\protect\label{f:svertka}
The dependence of binary WD coalescence rate (normalized to the 
present Galactic rate) on redshift $z$ for three representative
cosmological models: a standard one ($\Omega_{tot}=1,
\Omega_\Lambda=0$), 
a flat with cosmological constant
($\Omega_{tot}=1, \Omega_\Lambda=0.7$, and an open
without cosmological constant  
($\Omega_{tot}=0.2, \Omega_\Lambda=0$).}
\end{figure}

Fig. \ref{f:svertka} shows 
the dependence of the binary  white dwarf coalescence rate 
per a $10^{11} M_\odot$ galaxy on redshift calculated using Eq. 
(\ref{SFR}) in units of the (precisely unknown) 
galactic rate $\R_{wd}$ for different representative cosmological models:
the standard ($SCDM$: $\Omega_{tot}=1, \Omega_\Lambda=0$), a flat with 
the cosmological constant 
($\Lambda CDM$: $\Omega_{tot}=1,
\Omega_\Lambda=0.7$, and an open without the cosmological constant
($OCDM$: $\Omega_{tot}=0.2, \Omega_\Lambda=0$). 
The Green function G(t) for binary WD coalescence rate 
was taken from our numerical calculations
(Lipunov \& Postnov 1988; Lipunov et al. 1995; Lipunov, Postnov \&
Prokhorov 1996) which to a good approximation is 
\beq{wd_green}
G(t)=\frac{R_0}{1+k\frac{t-t(z_*)}{t_H}}\,, 
\eeq

where $t_H$ is the Hubble time, $k$ is a numerical factor,
$R_0$ is a constant calculated assuming that the present 
Galactic binary WD coalescence rate is
the mean value for a constant star formation:
\beq{green_norm}
\R=\frac{1}{t_H-t(z_*)}\int\limits_{t(z_*)}^{t_H}G(\tau)\,d\tau
\eeq

It is seen from this figure that the choice of the cosmological
model affects this function insignificantly.  

\begin{figure}
\centerline{  
\epsfxsize=0.8\hsize
%\epsfbox{../../rationew.eps}}
\epsfbox{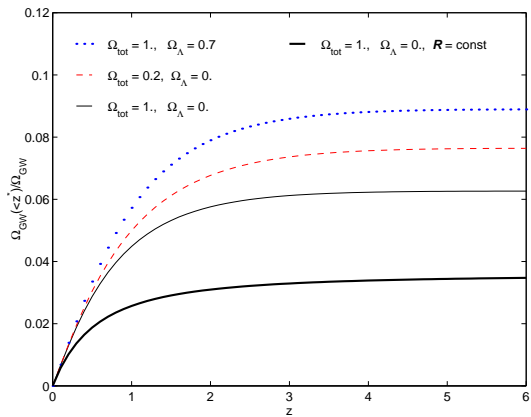}}
\caption{\protect\label{f:ratio}
The ratio of the energy density of the extragalactic 
stochastic GW background formed by merging binary WD  
$\Omega_{GW}(<z_*)$ to the mean Galactic level at frequency
$f=0.01$ Hz as a function of parameter $z_*$, the epoque of the inital 
star formation in the Universe. The global star formation 
rate history has the form 
(\protect\ref{SFR(z)}). Cosmological models are the same as in 
Fig.  \protect\ref{f:svertka}.}
\end{figure}

After having obtained $\R_G(z)$, we can easily calculate, 
substituting  $F(z)=\R_G(z)$ into Eq. (\ref{ratio}), the ratio  
$\Omega_{GW}(<z_*)/\Omega_{GW}(z=0)$ 
for different cosmological models.
The results are presented in Fig. \ref{f:ratio} for the same models as in 
the previous Figure. For comparison, the thick solid line 
shows the value $\Omega_{GW}(<z_*)/\Omega_{GW}$ 
assuming a constant star formation rate in the Universe
(Eq. [\ref{ratio_flat}]).

Our consideration 
of the stochastic GW background formation in the process
of extragalactic binary WD merging remains valid 
until redshifts where "boundary effects" of frequency 
distribution of binary stars are unimportant. In the frequency range 
$f_{obs}=10^{-3}-10^{-2}$ Hz this is always the case because the 
maximum proper frequency of the noise 
$f_{em}=f_{obs}(1+z)<10 f_{em}<0.1$ Hz for $z_*<6$.

\section{Conclusions}

Our results show the importance of taking into account effects of the global
star formation rate evolution. It is seen that in some realistic
cosmological models the stochastic background produced by extragalactic
merging binary WD can be about 0.1 of the mean galactic value. For a 1-year
LISA observation the mean (i.e. angle-averaged) Galactic background becomes
"transparent" at $f\sim 3\times 10^{-3}$ Hz, so at higher frequencies the
search for other GW backgrounds becomes possible. The level of the
extragalactic binary WD background (as well as any noise produced by
astrophysical sources) is proportional to the fraction of baryonic matter
$\Omega_b$, which lies in the range
$0.004<\Omega_bh_{100}^2<0.02$ (Fukugita et
al. 1997). Although most baryons are still in the form of ionized gas, we
can substitute, as an upper limit, the value $\Omega_bh_{100}^2=0.02$ into
Eq. (\ref{ratio}), thus increasing the extragalactic GW background by
four times. Such an extreme situation is feasible, for instance,
if all baryons had passed a stellar stage during an early star
formation burst at higher redshifts $z\simgt 6-10$, where spheroidal
systems formed rapidly (the so-called "third population stars"; e.g. the
model of Eggen, Lynden-Bell \& Sandage 1962). The evolutionary history of
the oldest ellipticals and low-surface brightness galaxy may also differ
significantly from the global averge star formation discussed above. The
traces of the early star formation is difficult to obtain by direct studies
of the UV luminosity density evolution because of a strong dust extinction.
Perhaps, the very detection of an isotropic GW background, together with
independent studies of far-IR background, would help revealing the true star
formation history at high redshifts.

At present, the lack of observational data on the SFR behaviour at high
redshift does not allow us to make more robust estimates. We conclude that
unless the global star formation rate continues increasing with redshift at
$z>3$, the extragalactic GW background energy density $\Omega_GW\sim
10^{-9}$ is ten times smaller than the mean galactic value
in the LISA frequency range $10^{-4}-10^{-1}$ Hz. 

The galactic binary GW noise will be modulated by the LISA orbital motion,
whereas the extragalactic one will not. If the latter is comparable with
galactic values at some galactic latitudes, it can impede detection by LISA
of some interesting relic cosmological GW backgrounds (e.g. Grishchuk 1997),
and specific statistical features of the relic GW should be used to separate
them against the noise from astrophysical sources.

\vskip\baselineskip

The authors 
acknowledge D. Hils and P.L. Bender for correcting an error in our
calculations.
The work was partially supported by Russian Fund for Basic Research
through Grant No 98-02-16801, the INTAS Grant No 93-3364, and 
by NTP "Astronomy" (project 1.4.4.1) of the Ministry of Science of Russia.


\begin{thebibliography}{99}

\bibitem{} Bender, P.L., Ciufolini, I., Danzmann, K. et al., 1996, 
LISA: Pre-Phase A Report (MPQ 208) (Max-Planck-Institut f\"ur Quantumoptil, 
Garching, Germany) 


\bibitem{} Bender, P.L.,  Hils, D.,  1997, Class. Quantum Grav., 14, 1439


\bibitem{BoyleTerlevich97} Boyle, B.J., Terlevich, R.J. 1997, MNRAS,
293, L49

\bibitem{Carroll_ea92} Carroll, S.M., Press, W.H., Turner, E.L. 1992, 
Ann. Rev. Astron. Astrophys., 30, 499


\bibitem{Connolly_ea97} Connolly, A.J., Szalay, A.Z., Dickinson, M. et al.
1997, ApJ, 486, L11


\bibitem{}Cutler, C., 1997, astro-ph/9703068

\bibitem{} Eggen, O.J., Lynden-Bell, D., Sandage, A.R. 1962, ApJ, 136, 748 

\bibitem{} Fukugita, M., Hogan, C.J., Peebles, P.J.E. 1996, Nature, 381, 489

\bibitem{} Fukugita, M., Hogan, C.J., Peebles, P.J.E. 1997, ApJ, in press
(astro-ph/9712020)


\bibitem{}Giampieri, G., Polnarev, A.G., 1997, MNRAS, 291, 149

\bibitem{Gri97} Grishchuk, L.P. 1997, Class. Quantum Grav., 14, 1445.


\bibitem{}Hils, D.L., Bender, P., Webbink, R.F. 1990, ApJ, 360, 75

\bibitem{Lilly_ea96} Lilly, S.J., Le Fevre, O., Hammer, F., Crampton, D. 1996, 
ApJ, 460, L1


\bibitem{LP87} Lipunov, V.M., Postnov, K.A. 1987, AZh, 64, 438

\bibitem{}Lipunov, V.M., Postnov, K.A.,   Prokhorov, M.E. 1987,
A\&A, 176, L1

\bibitem{}Lipunov, V.M., Nazin, S.N., Panchenko, I.E., Postnov, K.A., 
Prokhorov, M.E. 1995, A\&A, 298, 677

\bibitem{}Lipunov, V.M., Postnov, K.A.,  Prokhorov, M.E. 1996, 
Astrophys. Space Phys. Rev., Ed. by R.A.Sunyaev (Amsterdam: 
Harwood Acad. Publ.),
v. 9, p. 1

\bibitem{Loveday_ea92} Loveday, J., Peterson, B.A., Efstathiou, G., 
Maddox, S.J. 1992, ApJ, 390, 338


\bibitem{Madau_ea96} Madau, P., Ferguson, H.C., Dickinson, M. et al. 1996,
MNRAS, 283, 1388

\bibitem{Madau_ea97} Madau, P., Pozzetti, L., Dickinson, M. 1997,
ApJ, 498, 106


\bibitem{} Postnov, K.A., Prokhorov, M.E. 1998, ApJ, 494, 674


\bibitem{RosiZimmerman76} Rosi, S., Zimmermann, M. 1976,  
Astrophys. Space Sci., 45, 447


\bibitem{} Schutz, B.F. 1997, in Proc. 1977 Alpbach Summer School on 
Fundamental Physics in Space, ed. A Wilson, ESA Publ., in press 
(Preprint AEI 044, September 1997) 


\end{thebibliography}
\end{document}